\documentclass[onecolumn,showpacs,preprintnumbers,amsmath,amssymb]{revtex4}
\usepackage{CJK}
\usepackage{amssymb}
\usepackage{amsmath}
\usepackage{graphicx}
\usepackage{lscape}
\begin{document}
\begin{CJK*}{GB}{gbsn}
\title{The calculation of prompt fission neutron spectrum for $^{233}$U(n, f) reaction by the semi-empirical method}
\author{Yong-Jing Chen$^{1}$(³ÂÓÀ¾²)}
\email{cyj@ciae.ac.cn}
\author{JIA Min$^{2}$(¼ÖÃô)}
\author{Ting-Jin(ÁõÍ¢½ø)$^{1}$, SHU Neng-Chuan(ÊæÄÜ´¨)$^{1}$ }
\affiliation{$^{1}$ Department of Nuclear Physics, China Institute of Atomic Energy, Beijing 102413, China\\
$^{2}$2~(Department of Mathematics and Information Science, North China University of Water Conservancy and Electric Power, Zhengzhou 450011, China}

\begin{abstract}
The prompt fission neutron spectra for neutron-induced fission of $^{233}$U for low energy neutron (below 6 MeV)
are calculated using the nuclear evaporation theory with a semi-empirical method, in which the
partition of the total excitation energy between the fission fragments for the n$_{\textrm{th}}$+$^{233}$U
fission reactions are determined with the available experimental and evaluation data.
The calculated prompt fission neutron spectra agree well with the experimental data.
The proportions of high-energy outgoing neutrons of prompt fission neutron spectrum
versus incident neutron energies are investigated with the theoretical spectra,
and the results are consistent with the systematics.
We conclude that the semi-empirical method is sound,
and it could be a useful tool for prompt fission neutron spectra evaluation.
\end{abstract}

\pacs{25.85.Ec, 24.10.-i, 24.75.+i}

\maketitle

\section{Introduction}
The prompt fission neutron spectra and the 'sawtooth' data $\nu(A)$ of actinides are critical important nuclear
data for nuclear engineering and technologies, both in energy and non-energy applications.
Especially for $^{233}$U(n, f) reaction, because being the new generation nuclear fuel\citep{IAEA}, it attracts more and more attention.
The properties of prompt neutron are significant for the design of fusion-fission hybrid reactor,
thus a new calculation of these quantities with higher accuracy is required.
In addition, from a more fundamental point of view, studying the prompt fission neutron spectrum in detail can
provide valuable information on the understanding of the neutron induced fission process.

The early representations of the prompt fission neutron spectrum for actinides, in which many physical
effects were covered up, are the Maxwell and Watt spectrum representations with one or two parameters
that are adjusted to reproduce the experimental spectrum. The Los Alamos (LA) model\citep{Maland82} is one of the
most successful models for predicting the prompt fission neutron spectrum with an assumption of the
same triangular-shaped initial nuclear temperature distribution for both light and
heavy fragments, and was originally developed for $^{235}$U and $^{239}$Pu. Based on the LA model,
the multi-modal random neck-rupture model\citep{Brosa,Fan} has been applied to some calculations
of prompt neutron spectrum of several actinide nuclei isotopes by Ohsawa\citep{Ohsawa2000,Ohsawa2001}, Hambach\citep{Hambsch02,Hambsch03,Hambsch05}, Vladuca\citep{Vladuca01}, and Zheng\citep{Zheng} et al.
But most of these calculations are for $^{235,238}$U(n, f) and $^{239}$Pu(n, f) reactions,
only the calculation of Ref.\citep{Zheng} is for $^{233}$U(n, f) reaction,
but the nuclear temperature is still an assumption of simple triangular shaped.

In this article, we report the calculated results of the prompt fission  neutron
spectrum for neutron induced $^{233}$U fission reaction with a semi-empirical method,
which is very different from the LA model, such as the total excitation energy ($E_\textrm{TXE}$)
partition between the two complementary fragments, the nuclear temperature of each fragment and
the weight of the prompt fission neutron spectra for the light and heavy fragments.

In the present work, the information about the $E_\textrm{TXE}$ partition between the two complementary fragments
of n$_{\textrm{th}}$+$^{233}$U fission reaction are extracted from the available experimental and evaluation
data, and it is very important for the calculation of prompt fission neutron spectrum in the semi-empirical method.
With the Fermi-Gas model for nuclear energy level density and the initial excitation energy of every fission fragment,
the nuclear temperature of each fragment can be calculated, in which the constant temperature model are taken into account.
The spectrum in Center-of-Mass for each fragment are calculated by the semi-empirical method, and then are transformed to
the laboratory system. The calculated total spectra are synthesized by the chain yield and the prompt neutron number
and compared with the experimental data.

\section{Methodology}
\subsection{$E_\textrm{TXE}$ partition method}
We have used the semi-empirical method to describe the prompt fission neutron
spectrum and some other physical quantities of n+$^{235}$U fission reaction\citep{CPC2012}.
In this paper, this method has been applied to calculate the the prompt fission neutron
spectrum and the prompt fission neutron number for n+$^{233}$U fission reaction.
Let us briefly review this method.

In the semi-empirical method, the nuclear temperature of fission fragment is calculated by the
initial excitation energy of the fission fragment. But how to get the
initial excitation energy of each fission fragment from the available total excitation energy
is one of the long-standing problem about the nuclear fission
process. So it is important to know the partitioning of the total excitation energy ($E_\textrm{TXE}$)
between the two complementary fission fragments.
In the present work, we extract the $E_\textrm{TXE}$ partition
information from accumulated experimental and evaluated data for this reaction.

In the case of binary fission, the initial excitation energy of
each fragment is taken away by prompt neutrons and the prompt
$\gamma$ rays, and can be obtained by relevant experimental data.
For a pair of complementary fragments, if the initial excitation
energy of each fragment is obtained, then the energy partition
between the two complementary fragments can be deduced. For a
fission fragment $A_\textrm{L}$ or $A_\textrm{H}$, its initial
excitation energy can be expressed as:
\begin{equation}
E^*(A_\textrm{L,H})=\bar{\nu}_{\textrm{exp}}(A_\textrm{L,H})\langle\eta\rangle(A_\textrm{L,H})+\bar{E}_{\textrm{exp},\gamma}(A_\textrm{L,H}).\label{eq1}
\end{equation}
Where $\langle\eta\rangle$ is the average energy removed by an
emitted neutron from fragment $A_\textrm{L}$ or $A_\textrm{H}$,
and it is composed of the average neutron kinetic energy
$\varepsilon_{\textrm{exp}}(i)$ and the neutron separation energy
$S_{\textrm{n}}(i)$ ( i stands for $A_\textrm{L}$ or $A_\textrm{H}$).
The sum of $E^*(A_\textrm{L})$ and
$E^*(A_\textrm{H})$ is the total excitation energy $E_\textrm{TXE}$.
Using Eq.(\ref{eq1}), the ratio $R(A_\textrm{L,H})$ of
$E^*(A_\textrm{L,H})$ with respect to $E_\textrm{TXE}$, which shows the $E_\textrm{TXE}$
partition between the two fragments, can be expressed as follows with the experimental data:
\begin{eqnarray}
R(A_\textrm{L,H})&=&\frac{E^*(A_\textrm{L,H})}{E_\textrm{TXE}}   \nonumber  \\
&=&\frac{\bar{\nu}_{\textrm{exp}}(A_\textrm{L,H})\langle\eta\rangle(A_\textrm{L,H})+\bar{E}_{\textrm{exp},\gamma}(A_\textrm{L,H})}{\sum\limits_    {i=A_\textrm{L},A_\textrm{H}}[\bar{\nu}_{\textrm{exp}}(i)\langle\eta\rangle(i)+\bar{E}_{\textrm{exp},\gamma}(i)]}.
\label{eq4}
\end{eqnarray}
For the thermal neutron-induced fission reaction of $^{233}$U, the
relevant experimental data are taken from Ref.\citep{NST}.
In this work, all quantities entering the
calculation are replaced by the evaluated data
$\bar{\nu}_{\textrm{eval}}(A)$, $\varepsilon_{\textrm{eval}}(A)$
and $\bar{E}_{\textrm{eval},\gamma}(A)$, respectively. These
values are obtained by fitting the experimental data or by
interpolation and extrapolation when no experimental data are
available.

Based on the evaluated data, the energy partition between two
complementary fragments of n$_{\textrm{th}}$+$^{233}$U fission
reaction are calculated according to Eq.(2). The results
are shown in Fig. 1 by the short dot line, and the solid line is the smoothed results.
This result is very similar to the case of n+$^{235}$U
fission reaction, and a deep valley appears around  A $\sim$ 130. The minimum close
to A $\sim$ 130 is due to the shell closures N $=$ 82, Z $=$ 50 that lead to spherical fission fragments.

\begin{figure}[htbp]
\includegraphics[height=2.in,width=3in]{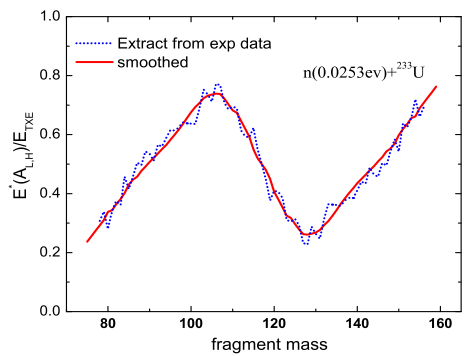}
\caption{(Color online) The energy partition between two fragments in the thermal neutron induced
$^{233}$U fission reaction.}
\end{figure}

At moment, except for the thermal neutrons, there are no enough experimental
data available for other neutron energy induced $^{233}$U fission, so how to get the $E_\textrm{TXE}$ partition
for other incident neutron energy ($R_{\textrm{E}_{\textrm{n}}}(A)$) is a critical problem.
We had discussed this problem for n+$^{235}$U fission
reaction with a semi-empirical method, and some technical details concerning this method are given in Ref.\citep{cpc2010}.
In this paper, we adopt the similar method to get the $E_\textrm{TXE}$ partition dependence of the incident
neutron energies (below 6 MeV) for n+$^{233}$U fission reaction,
and the systematic parameters of fission fragment mass
distribution of n+$^{233}$U fission system are taken from Ref.\citep{interreport2012}.

For a binary fission reaction, the total excitation energy $E_\textrm{{TXE}}$ of fission
fragment pair is given as follows:
\begin{eqnarray}
E_\textrm{{TXE}}=E^{*}_{\textrm{r}}(A_\textrm{L}+A_\textrm{H})+B_\textrm{n}(A_\textrm{c})+E_\textrm{n}-E_{\textrm{TKE}}(A_\textrm{L}+A_\textrm{H}),
\end{eqnarray}
where $E^{*}_{\textrm{r}}(A_\textrm{L}+A_\textrm{H})$ is the energy
released in the fission process, which is given by the difference between the compound nucleus
and the FF masses. $B_\textrm{n}(A_\textrm{c})$ is the neutron binding energy of the fission
compound nucleus. $E_\textrm{n}$ is the kinetic energy of the neutron inducing fission.
$E_{\textrm{TKE}}(A_\textrm{L}+A_\textrm{H})$ is the total kinetic energy of both light
and heavy fragments. For the n$_{\textrm{}}$+$^{233}$U fission reaction, the
initial excitation energy of each fragment, $E^*(A)$, can be obtained by means of the energy
partition $R_{\textrm{E}_{\textrm{n}}}(A_{})$ and the total excitation energy $E_\textrm{{TXE}}$:
\begin{equation}
 E^{*}(A_{})=R_{\textrm{E}_{\textrm{n}}}(A_{})\times E^{}_{\textrm{TXE}}.
\end{equation}

\subsection{Neutron evaporation}

At higher nuclear excitation energies, within the Fermi gas model for nuclear energy level density, respectively.
The initial fission fragment
energy $E^{*}(A)$ is simply related to the nuclear temperature $T$ as the following:
\begin{equation}
T_{}=\sqrt{\frac{E^{*}(A)-S_\textrm{n}(A)}{a_\textrm{A-1}}},
\end{equation}
where $a_{A-1}$ and $S_n$ are the level density parameter and the neutron separation, respectively.  At lower excitation energies, we assumed a constant
temperature regime for neutron evaporation. The probability for the fission fragment to emit a neutron at
a given kinetic energy is obtained by Weisskopf spectrum at this particular temperature\citep{WeiSKPF37}.
Assuming a constant value of the cross section of inverse process of compound nucleus formation,
the normalized prompt fission neutron spectrum $\phi(\varepsilon)$ in the center of mass system is
\begin{equation}
\phi(A,T,\varepsilon)=\frac{\varepsilon}{T_{}^2}\textrm{exp}(-\varepsilon/T)],
\label{eq3}
\end{equation}
where $\varepsilon$ is the center-of-mass neutron energy.

For a fragment with excitation energy $E^{*}(A)$, it could de-excite through emitting neutrons and $\gamma$ rays.
The excitation energy of fragment will decrease after a neutron is emitted from a fragment,
this will decrease the nuclear temperature $T$ as well.
The prompt fission neutron spectra at different temperature $T_i$ were calculated by using the Eq.(5) for each fragment.
The total prompt fission neutron spectrum in the Center-of-Mass of every fragment
is written as $\phi(A,\varepsilon)$ and can be obtained
by summing all of them up with the corresponding weight $P^{''}_\textrm{N}(i)$:

\begin{equation}
\phi(A,\varepsilon)=\sum\limits_{i}\frac{\varepsilon}{T_{i}^2}\textrm{exp}(-\varepsilon/T_{i})\times P^{''}_\textrm{N}(i).
\label{eq8}
\end{equation}
$P^{''}_\textrm{N}(i)$ is the number of emitting the i-th neutron for emitting the total N neutrons.
For a given fragment $A$, the sum of $P^{''}_\textrm{N}(i)(i=1,N)$ is equal to $\bar{\nu}(A)$.
The detail discussion about $P^{''}_\textrm{N}(i)$ can refer to Ref.\citep{CPC2012}.

Given the center-of-mass neutron energy spectra of every fragment, the neutron energy spectra $\Phi(A,E)$
in the laboratory system can be obtained by assuming that neutrons are emitted isotropically in the center
of mass frame of a fission fragment. The total prompt fission neutron spectra of all fragments can be expressed as:
\begin{equation}
N(E)=\sum\limits_{j}Y(A_{j})\bar{\nu}(A_{j})\Phi(A_{j},E),
\label{eq9}
\end{equation}
where $j$ stands for all fission fragments. $Y(A)$ is the chain yield, and $\bar{\nu}(A)$ is the average prompt fission neutron number
as a function of the fission fragment mass number, which also has been calculated in this work.

\section{Results and discussions}
The experimental data of prompt fission neutron spectrum for n+$^{233}$U fission reaction is scarce,
and there are only a few data for incident thermal neutron and 0.55 MeV can be found
from the open published references. Figure 2 shows the calculated prompt fission neutron spectra for two energy points
and compared with the experimental data as well as the Maxwell spectra. Here, the mean laboratory neutron energy of Maxwell spectrum
is equal to that given by the calculated theoretical spectrum. In Fig. 2, the solid curves indicate the calculated results,
the dash lines show the Maxwell spectra, and the other symbols are the experimental data taken from International Experimental
Neutron Data Library EXFOR\citep{exfor2012}. It is clear from Fig. 2 that the calculated spectra are much better agree
with the experimental data than the Maxwell spectra.
For thermal case, the calculated spectra are in good agreement
with the most experiment data covering different energy ranges except for the experimental data of B.I.Starostov(1985) above 5 MeV,
but in fact their two sets of data are discrepant with each other.
For 0.55 MeV case, we note that the agreement between the present calculation and the experimental data is good,
but the calculated spectrum appears to be somewhat harder from 6.0 MeV to 9.0 MeV.
Moreover we calculated the data for the entire energy range (0-20MeV) required in evaluations. Unfortunately,
there are no experimental data above 12 MeV for these fission spectra,
so it is not possible to determine the agreement is good or not at the tail region of the spectrum.

\begin{figure}[htbp]
\includegraphics[height=5.in,width=6in]{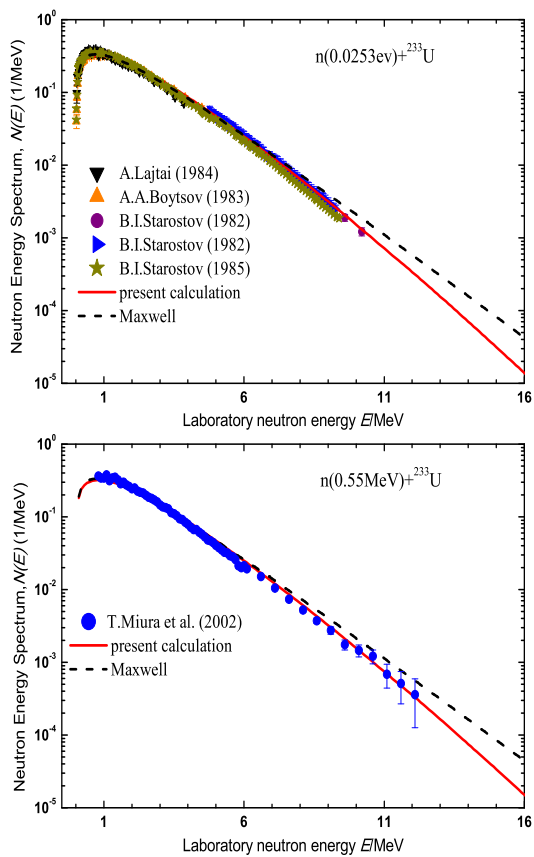}
\caption{(Color online)The total prompt fission neutron spectra for thermal neutron and 0.55 MeV compared
with the experimental data\cite{exfor2012} and the Maxwell spectra for n+$^{233}$U fission reaction.}
\end{figure}

The good agreement between experiment and calculation in Fig. 2 shows that the method used in this
work for $^{233}$U(n, f) reaction is valid. For other energies below 6 MeV,
though no experimental data are available, the calculated prompt fission neutron spectra are also presented in Fig. 3,
and the incident neutron energies are presented in figure.
It is clear from this figure that with increasing of incident neutron energy,
the spectra are generally hard in the tail region and soft in the low-energy region.
This tendency is reasonable, because increasing the incident neutron energies increases the
total excitation energy ($E_\textrm{TXE}$), and increases the nuclear temperature.

\begin{figure}[htbp]
\includegraphics[height=2.5in,width=3.5in]{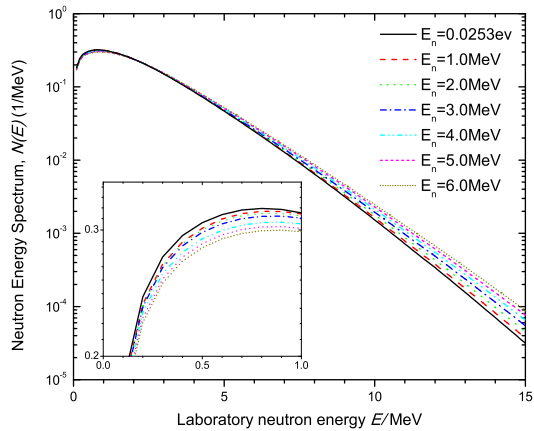}
\caption{(Color online)The total prompt fission neutron spectra versus incident neutron energies.}
\end{figure}

With these theoretical spectra of Fig. 3, we can investigate the proportion of high-energy outgoing neutron of
total prompt fission neutron spectrum versus incident neutron energies for n+$^{233}$U fission reaction.
Generally, theoretical prompt fission neutron spectra are normalized to unity when integrated from zero to infinity.
So the integral of the theoretical spectrum over an arbitrary energy interval can give us
the proportion of outgoing neutron in considering energy range.
Figure 4 shows us proportions of high-energy outgoing neutron from from 1.9\% to 2.7\%
as a function of incident energies from 0.0253ev to 6 MeV,
and the high-energy outgoing neutron energy range considered in this work is 7 MeV to 20 MeV.
It is clear from Fig. 4 that the proportions of high-energy outgoing neutron increase
with increasing of the incident neutron energy.
These results also can be seen directly from Fig. 3. The incident neutron energy increase,
the tail region becomes harder, namely, the number of high-energy outgoing neutron increases.
Therefore, it can be predicted that the component of high-energy outgoing neutrons will become
more with increasing of incident neutron energy. So the high-energy outgoing neutrons of
prompt fission neutron spectrum attract more attention in various the nuclear energy applications.

\begin{figure}[htbp]
\includegraphics[height=2.5in,width=3.5in]{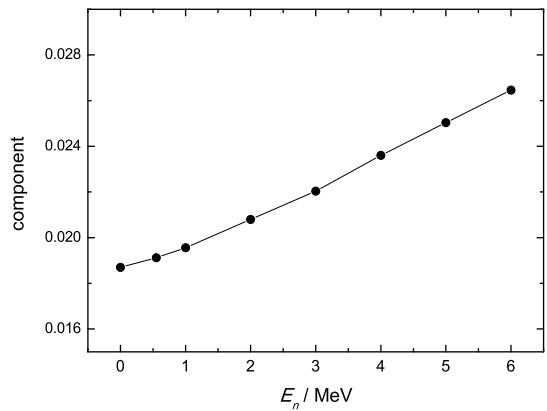}
\caption{(Color online)The proportion of high-energy outgoing neutron as a
function of incident neutron energies for $^{233}$U(n,f) reaction.}
\end{figure}

Finally, the prompt fission neutron multiplicity for $^{233}$U(n, f) reaction
as a function of incident neutron energy were also calculated.
Figure 5 shows us the total average prompt fission neutron multiplicity
$\bar{\nu}$ and compared with the evaluated data.
The calculated results agree well with the evaluated results at $E_n$ below 6 MeV.

\begin{figure}[htbp]
\includegraphics[height=2.5in,width=3.5in]{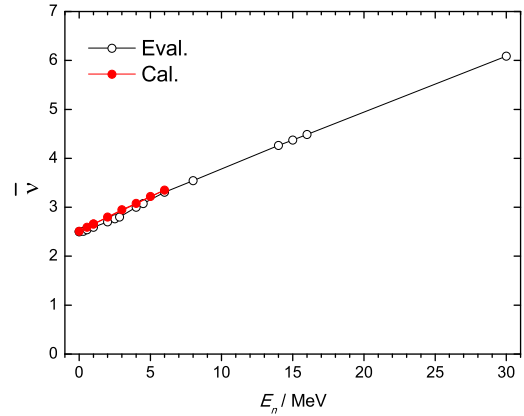}
\caption{(Color online)The total prompt neutron number as a function of incident neutron energies.}
\end{figure}

\section{Summary}
With the available experimental and evaluated data, the energy partition of the
total excitation energy between the fission fragments for the n$_{\textrm{th}}$+$^{233}$U
fission reactions is investigated. The prompt fission neutron spectra for neutron-induced fission of $^{233}$U at
incident neutron energy $E_n=$ 0.0253 ev and 0.55 MeV are calculated with a semi-empirical method,
and the calculated results are in good agreement with the experimental prompt fission neutron spectra.
This indicates that the semi-empirical method is sound.
We also present additional calculations of the prompt fission neutron spectrum for
neutron-induced fission of $^{233}$U at other energies below 6 MeV. With these theoretical spectra,
we discussed the proportions of high-energy outgoing neutrons of
prompt fission neutron spectrum versus incident neutron energies,
and the results are consistent with the theoretical systematics.
In summarizing, the semi-empirical method has described the prompt fission neutron spectra of neutron-induced $^{233}$U and
$^{235}$U fission reactions very well. So it could be a useful tool for evaluation of prompt fission neutron spectra,
and can be applied on other actinide neutron induced fission.

\end{CJK*}

\section{Acknowledgments}
The work were supported by National Natural Science Foundation of China (11205246,91126010,91226102).

\end{document}